\documentclass[sigplan,nonacm,10pt]{acmart}
\setcopyright{none}

\usepackage{svg}
\usepackage{float}

\author{Konstantinos Mores}
\affiliation{
    \institution{NTUA, Greece}
}
\author{Stratos Psomadakis}  
\affiliation{
    \institution{NTUA, Greece}
}
\author{Georgios Goumas}  
\affiliation{
    \institution{NTUA, Greece}
}

\begin{document}

\title{eBPF-mm: Userspace-guided memory management in Linux with eBPF}

\maketitle

\section*{Research Problem}
As datacenter workloads become ever more data intensive, the Operating System (OS) memory management (MM) subsystem faces increasing pressure.
Additionally, the growing chasm between processor and memory speed has made memory accesses the main bottleneck for many modern applications~\cite{memwall}. 
Virtual-to-physical address translation (AT) contributes substantially to memory access latency~\cite{ATWall}.
CPUs employ hardware caches, Translation Lookaside Buffers (TLBs), to speed up this operation.
To increase the TLB reach and  alleviate the AT overhead, OS and HW cooperatively implement and support huge pages \cite{largepagesurvey}.
Huge pages are virtually and physically contiguous memory areas, larger than 4KiB, that can be translated using a single TLB entry. 

Despite the promising performance benefits of huge pages, they seem to underperform in modern OSes~\cite{largepagesnumaharmful} leading to many applications  to advise disabling them \cite{mongodisablethp}.
This can be partially attributed to the cost-obliviousness of many MM huge-page policies.
For example, Linux Transparent Huge Pages (THP)~\cite{thp} try to greedily allocate a huge page every time a memory area is first touched, not considering that this area might be underutilized or that it does not benefit from being backed by huge pages.

However, setting up a huge page is not cost-free. 
The kernel first needs to find an available, properly aligned physical memory region. 
This process is inexpensive if there is an abundance of physical memory, but  on a fragmented machine this could trigger costly operations, like compaction.
The huge page must then be prepared. 
In the case of file-backed memory this means fetching the data from the disk, and in the case of anonymous mappings it means zeroing the contents of the huge page. 
Therefore, if the costs of setting up a huge page outweigh its benefits not only the current process takes longer, but the whole system is susceptible to more slowdowns due to fragmentation~\cite{cbmm}.
Consequently, the kernel should take great care deciding whether or not to use huge pages to back application memory.

The x86 architecture supports 2 huge pages sizes, i.e., 2MiB and 1GiB.
ARMv8-A and RISC-V extend the supported huge page sizes, using a similar mechanism which utilizes unused bits of the page table entries to designate 
contiguous groups of pages~\cite{ArmManual,hugetlbfscontig,riscv-napot,ncpapad-svnapot}. 
This further complicates the OS huge-page policies, by adding 
64KiB and 32MiB huge pages to the mix.
In order to support these new huge page sizes, Linux switched to a mutli-sized THP mechanism \cite{mthp,arm-patch}.
However, mTHP doesn't allow for fine-grained control over which applications or which parts of the address space of an application should be backed by which size.

We argue that the OS memory manager needs to be more flexible in order to navigate the trade-offs of using different huge page sizes for different applications.

\section*{Our contributions}
eBPF~\cite{ebpfio, ebpfdocs, bgebpf, lzrebpf} is a revolutionary kernel technology that allows custom-written user programs to be attached to various points in the kernel (hooks) and be executed in kernelspace when the corresponding event is triggered. 
The programs’ security is ensured through an in-kernel verifier \cite{ebpfverifier} at load time. 
eBPF makes the kernel programmable without the need for kernel modules or rebooting.
    
We leverage eBPF in order to implement custom policies in the Linux memory subsystem. 
Inspired by CBMM~\cite{cbmm}, we create a mechanism that provides the kernel with hints regarding the benefit of promoting a page to a specific size.
We introduce a new hook point in Linux page fault handling path for eBPF programs, providing them the necessary context to determine the page size to be used.

We then develop a framework that allows users to define profiles for their applications and load them into the kernel. 
A profile consists of memory regions of interest and their expected benefit from being backed by pages of 64KB, 2MB and 32MB. In our evaluation, we profiled our workloads to identify hot memory regions using DAMON~\cite{damon,damon-linux}. 

\begin{figure}[h]
\caption{eBPF-mm workflow}
\label{ebpfmm}
\includegraphics[width=\columnwidth]{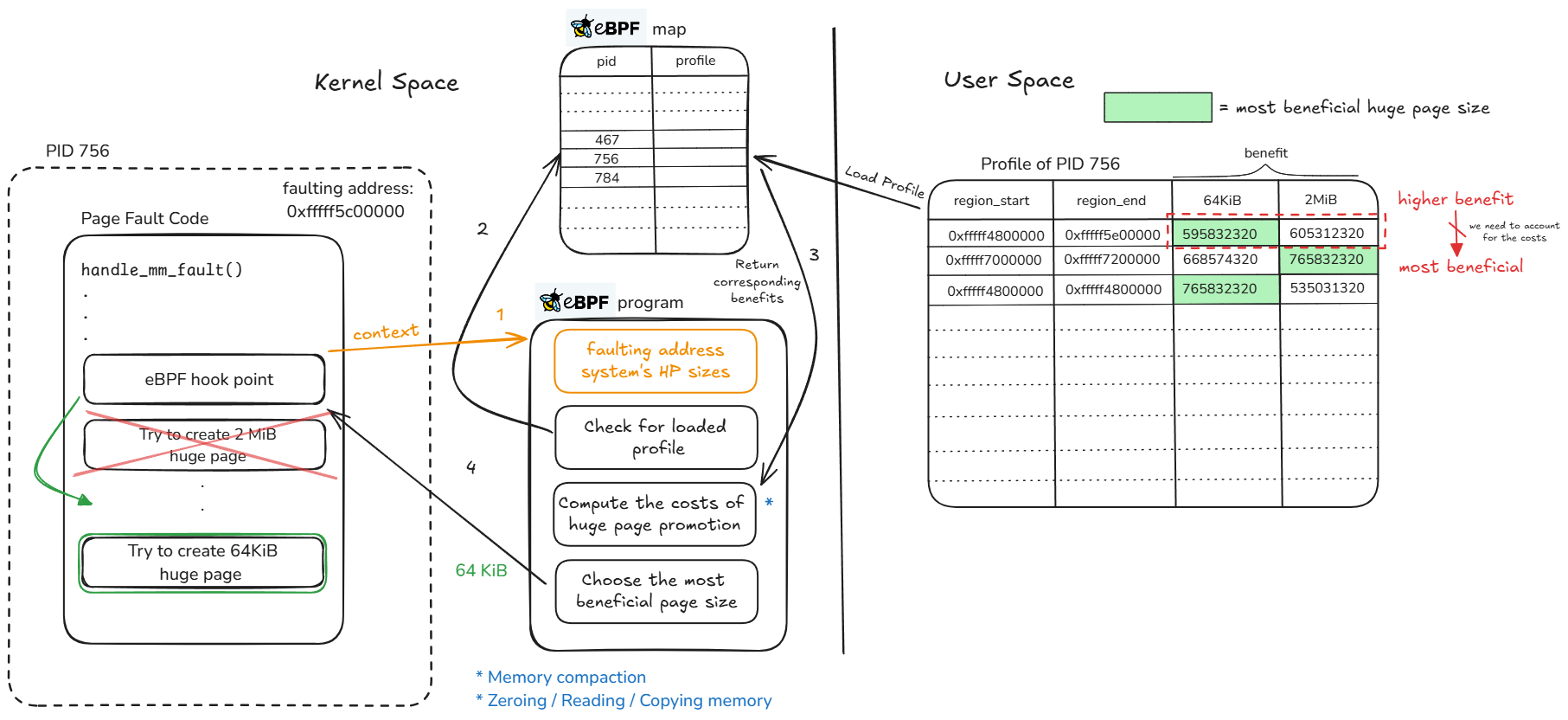}
\end{figure}

The user-provided profiles are consumed by eBPF programs attached to the aforementioned hooks. We now describe our proposed eBPF program in detail (Fig.~\ref{ebpfmm}). 
A page fault occurs which triggers our eBPF program doing the following:
\begin{itemize}
\item Check if the faulting process has a loaded profile. 
\item Search the profile for a profiled memory region containing the faulting address.
\item Compute the cost of promoting a huge page from real time system data.
\item Choose the most beneficial page size for the fault. 
\end{itemize}

To estimate the promotion cost, we assume that the primary contributors to this cost are the time required to prepare a huge page (zeroing) and the time needed to locate an available one (compaction). 
We empirically calculate a fixed cost for both zeroing and, in case of fragmentation, compaction.
To estimate the promotion benefit, we monitor the address space of a process with DAMON in the granularity of the huge page size and use the reported access frequency as a proxy. 

\begin{figure}[h]
\caption{eBPF-mm preliminary results for astar}
\label{ebpfmm-res}
\includegraphics[width=\columnwidth]{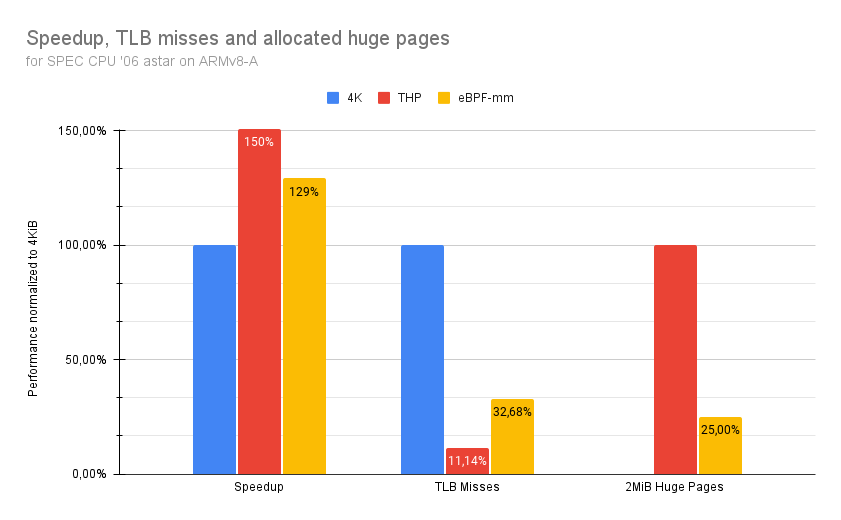}
\end{figure}

Fig.~\ref{ebpfmm-res} shows some preliminary results for the astar benchmark from the SPEC CPU 2006 suite~\cite{spec06}. Userspace huge page size hinting via eBPF-mm is able to provide competitive performance to THP while using only a fraction of 2MiB. eBPF-mm achieves this by only backing the AT-intensive parts of astar's address space with 2MiB huge pages.

\vspace*{-5mm}

Our system requires few modifications to the kernel, while also preserving the default behaviour of Linux. 
If a program is not attached to the hook or if the faulting process does not have any loaded profile,
Linux will not deviate from the standard code path, thus imposing zero overhead on non-hinted faults.

We intend to extend our eBPF mechanism to also support asynchronous huge page promotions (via the khugepaged kernel thread). We're also considering to incorporate more eBPF hooks in the Linux memory manager, in order to enable flexible userspace-driven policies, not only for huge page size selection but also for other memory management aspects, such as memory reclamation, when the system is under memory pressure, and page placement for memory tiering~\cite{memtis}.

\newpage

\bibliographystyle{IEEEtranN}
\bibliography{references}

\end{document}